# Patents as Knowledge Artifacts: An Information Science Perspective on Global Innovation.

M. S. Rajeevan[1], B. Mini Devi[2]

[1]Department of Library and Information Science, University of Kerala, Trivandrum, Kerala, India. ORCID: 0009-0009-3320-3078, Email: rajeevanms_2025@keralauniversity.ac.in

[2]Department of Library and Information Science, University of Kerala, Palayam, Trivandrum, Kerala, India. ORCID: 0000-0003-2736-7999, Email: drminidevi1968@gmail.com

*Abstract*

*In an age of fast-paced technological change, patents have evolved into not only legal mechanisms of intellectual property, but also structured storage containers of knowledge full of metadata, categories, and formal innovation. This chapter proposes to reframe patents in the context of information science, by focusing on patents as knowledge artifacts, and by seeing patents as fundamentally tied to the global movement of scientific and technological knowledge. With a focus on three areas, the inventions of AIs, biotech patents, and international competition with patents, this work considers how new technologies are challenging traditional notions of inventorship, access, and moral accountability.The chapter provides a critical analysis of AI's implications for patent authorship and prior art searches, ownership issues arising from proprietary claims in biotechnology to ethical dilemmas, and the problem of using patents for strategic advantage in a global context of innovation competition. In this analysis, the chapter identified the importance of organizing information, creating metadata standards about originality, implementing retrieval systems to access previous works, and ethical contemplation about patenting unseen relationships in innovation ecosystems. Ultimately, the chapter called for a collaborative, transparent, and ethically-based approach in managing knowledge in the patenting environment highlighting the role for information professionals and policy to contribute to access equity in innovation.*

1. Introduction

In today's knowledge economy, patents are not just legal instruments for protection of intellectual property; they are organized sources of technological information. They contain detailed information about inventions including technical details, methods, and possible uses, positioning them as crucial points of distribution of scientific and technological knowledge world-wide[1].From the viewpoint of information science, patents can be thought of as "knowledge artifacts" documents that systematically categorize and codify invention. Patents are part of the information infrastructure of contemporary innovation systems for the transfer of knowledge across disciplines and borders[2].

---

[1] COMMITTEE ON ADVANCING COMMERCIALIZATION FROM THE FEDERAL LABORATORIES ET AL., ADVANCING COMMERCIALIZATION OF DIGITAL PRODUCTS FROM FEDERAL LABORATORIES (2021), https://www.nap.edu/catalog/26006 (last visited Apr 24, 2025).

The emergence of artificial intelligence (AI) has complicated the patent environment. AI-mediated inventions complicate traditional understandings of inventorship, as most legal frameworks correctly recognize only natural persons as inventors. The United States Patent and Trademark Office (USPTO) has provided guidance stating that while AI may assist in the invention process, a human must have made a significant contribution to be considered an inventor[3]. With respect to biotechnology, ethical issues intertwine with patent law. Ethical licensing allows patent owners to impose conditions on licenses based on ethical considerations. The relationship between innovation, ethics, and intellectual property is complicated[4]. The use of patents worldwide relates to larger historical patterns of innovation and competition. Patent applications can show the direction of technological advancement and show an indication of how nations are configured in the global innovation system[5].

In this chapter, we will examine the complex role of patents as knowledge artifacts in information science. By looking at the complexity of AI-invented inventions, ethical challenges posed by biotech patents, and the global competition over patents, we want to figure out how information is created, value managed, and distributed in the context of modern innovation.

## 2. Theoretical Lens: Information Science and Knowledge Artifacts

Information science ultimately pertains to the processes by which knowledge and information are created, organized, stored, retrieved, and used. As society moves further into the knowledge economy, the boundaries between legal instruments, scientific documentation, and information systems continue to dissolve. Patents serve as one of the most explicit examples of the convergence of legal instruments, scientific knowledge, and structured knowledge. It operates as a legal instrument and organized body of knowledge.

Patents are knowledge artifacts from an information science view; codified representations of human thought and innovation, represented as part of formal systems of classification (or indexing) and access. Information can be considered a process or knowledge, and can also be considered a thing, either in a physical form or a digital object, that can embody knowledge used to communicate and recreate ideas[6].

---

[2] Patents as Scientific Information 1895-2020 (PASSIM), https://liu.se/en/research/passim (last visited Apr 24, 2025).
[3] Inventorship Guidance for AI-Assisted Inventions, FEDERAL REGISTER (2024), https://www.federalregister.gov/documents/2024/02/13/2024-02623/inventorship-guidance-for-ai-assisted-inventions (last visited Apr 24, 2025).
[4] Stanford Law School, *Using Patents as a Gavel: Governing Biotechnology With Ethical Licensing Restrictions*, STANFORD LAW SCHOOL (2024), https://law.stanford.edu/2024/10/14/using-patents-as-a-gavel-governing-biotechnology-using-ethical-licensing-restrictions/ (last visited Apr 24, 2025).
[5] National Science Board, NATIONAL SCIENCE BOARD, http://www.nsf.gov/nsb (last visited Apr 24, 2025).
[6] Michael K. Buckland, *Information as Thing*, 42 JOURNAL OF THE AMERICAN SOCIETY FOR INFORMATION SCIENCE 351 (1991), https://onlinelibrary.wiley.com/doi/abs/10.1002/%28SICI%291097-4571%28199106%2942%3A5%3C351%3A%3AAID-ASI5%3E3.0.CO%3B2-3 (last visited Apr 24, 2025).

## 2.1 Patents as Structured Information Objects

Each patent is a highly standardized document, with elaborate technical descriptions, illustrations, metadata regarding inventors and assignees, and claims - the legal boundaries of an invention. These features serve the basis for systematic indexing using classification systems such as the International Patent Classification (IPC) and the Cooperative Patent Classification (CPC), in the organization of huge amounts of technical knowledge[7]. The standardization in the structure of a patent helps with precise retrieval, comparison, or citation, similar to scholarly literature.

Bibliographic databases guide researchers through scholarly publications in the same way patent databases (such as Espacenet, WIPO Patentscope, USPTO) help with information retrieval pertaining to technology and innovation. These systems exhibit characteristics that depend on tools, like controlled vocabularies, ontologies, and the development of schemas for metadata, that have their origins in library and information science.

## 2.2 Knowledge Artifacts and Innovation Systems

Patents serve as boundary objects in innovation systems; they link technical, legal, economic, and policy domains simultaneously, used by scientists, engineers, patent examiners, legal experts, policy makers, and business strategists. To Star and Griesemer (1989), boundary objects are flexible enough to accommodate different perspectives, while robust enough to maintain identity across them[8]. Patents function as boundary objects, allowing disparate stakeholders to not only collaborate but also compete, while sharing the same information.

In addition, patents promote knowledge dissemination through mandatory disclosure. In exchange for exclusive rights, inventors must disclose their inventions to the public. In other words, inventors must provide detailed descriptions so that the inventions become publicly accessible. This principle of disclosure is a fundamental aspect of the information function of patents, enabling further research and development, even if legal rights may restrict certain ways to use the knowledge.

## 2.3 Information Ethics and Access

Information science equally provides the ethical lens for how we talk about access to and control over patent information. The issues of information asymmetries, digital divides, and equitable access to scientific knowledge are vital, especially when patents speak to the access to important medicines, genetic resources, or climate technologies. Open data, public domain, and commons-based knowledge sharing contradict traditional intellectual property frameworks and require systems for broadening access to new knowledge[9].

---

[7] International Patent Classification (IPC), CLASSIFICATION-IPC, https://www.wipo.int/web/classification-ipc (last visited Apr 24, 2025).

[8] Susan Leigh Star & James R. Griesemer, *Institutional Ecology, `Translations' and Boundary Objects: Amateurs and Professionals in Berkeley's Museum of Vertebrate Zoology, 1907-39*, 19 SOC STUD SCI 387 (1989), https://doi.org/10.1177/030631289019003001 (last visited Apr 24, 2025).

[9] UNDERSTANDING KNOWLEDGE AS A COMMONS: FROM THEORY TO PRACTICE, https://direct.mit.edu/books/edited-volume/3807/Understanding-Knowledge-as-a-CommonsFrom-Theory-to (last visited Apr 24, 2025).

## 3. AI-Generated Inventions and the Information Problem of Authorship

The rapid emergence of artificial intelligence (AI) in research and development has disrupted the way inventions are created, documented, and claimed. AI systems can independently derive and devise solutions to complex problems, synthesize chemical compounds, and design new products with little human input. This emergence brings added complexities to the traditional principles of authorship and inventorship in patent systems.

### 3.1 The Challenge to Traditional Inventorship Models

In current patent law frameworks around the world, including the United States, Europe and India, an inventor must be, by definition, a natural person. Patent law posits that inventive acts involve mental acts, creativity, and purposeful intent involving human beings. But AI tools are already doing what human inventors do, including a specific tool called DeepMind's AlphaFold and other generative design tools in engineering - sometimes without direct human input, and sometimes without foreknowledge of the results[10].

This reveals what we could refer to as the author attribution information problem: how do we attribute credit and credit for the creative acts of partially or entirely delegated by non-human agent? Although AI does not have legal personhood, its output is becoming embedded in patent applications raising issues of transparency, traceability, and accountability in the innovation record.

### 3.2 Patent Documentation and Information Traceability

From the point of view of information science, the matter is one of legality and, more importantly, epistemology and archiving. Patents are expected to be documents that trace human creativity-- who did what, and importantly, who, when, and how. With AI inventions, tracing the provenance of ideas is murky. This is complicated by the "black box" characteristic of many AI models, whose decisions are not always interpretable or replicable, which challenges full disclosure, a necessary feature of patent law[11].

Furthermore, there is no standard yet for how to document or cite the role of AI in the inventive process. Should the AI be listed as a tool, a co-creator, or simply omitted? These questions affect the clarity and retrievability of information in patent databases, and consequently, the future utility of such patents as knowledge artifacts.

### 3.3 Information Ethics and Recognition

This also raises issues of information ethics: if AI systems facilitate socially or commercially valuable inventions, who gets credit? Should inventors disclose how much AI contributed as part of their obligation to provide transparency and equitable access to knowledge?

---

[10] Introduction: Artificial Intelligence and the Law, , *in* THE REASONABLE ROBOT: ARTIFICIAL INTELLIGENCE AND THE LAW 1 (Ryan Abbott ed., 2020), https://www.cambridge.org/core/books/reasonable-robot/introduction-artificial-intelligence-and-the-law/DB801A3932CB86F1E08A6A10ACC91A8A (last visited Apr 24, 2025).

[11] What Is Black Box AI and How Does It Work? | IBM, (2024), https://www.ibm.com/think/topics/black-box-ai (last visited Apr 24, 2025).

Recently, policies have started to address these questions. For instance, the United States Patent and Trademark Office (USPTO) published guidance that acknowledged AI can help an inventor, but only a human can be listed as an inventor, reaffirming that patent attribution is human-centered[12].

However, this perspective may have to continue to shift as A.I. tools become more autonomous. From an information science perspective, that means falling into the urgency of needing new metadata, attribution, and archival policies to facilitate non-human creators without compromising fundamental concepts of disclosure and accountability.

4. **Biotech Patents and Ethical-Information Dilemmas**

Biotechnology has been in the gray space between innovation, ethics, and regulation for a long time. Making advancements in biotechnology can raise serious questions about what can be legitimately patented (and what should be). Whether through genetically modified organisms (GMOs) or CRISPR gene-editing technologies, biotechnological advances focus on substantial issues not only of what is patentable but also what could/should be patented. In essence, from an information science route, this space lays out the ethical problems of access, control, and disclosure in the patenting systems of essential knowledge.

**4.1 Patents on Life and the Moral Limits of Codification**

Many biotech inventions are living things, genetic sequences, or biological products. When patenting these inventions is performed, life becomes a coded information object to store, communicate, and monetize—the abstracted biological complexity into claims, sequences, and classifications. While this allows for the storage, communication, and monetization of a clear innovated entity, there are serious concerns of commodification of nature itself and the ownership of the world's genetic commons[13].

The ethical discussion becomes even more severe when the patents involve essential medicines, diagnostic equipment, or agriculture innovations related to food security. We as information science professionals should think not only about the representation of that information but also about its societal function—who gets it, on what basis, and for what purposes.

**4.2 Ethical Licensing and Information Access**

Some organizations have looked to ethical licensing models to address these worries. Ethical licensing is a framework whereby patents are licensed under terms that restrict the use of a particular licensed invention for unethical purposes or grant access under certain conditions to support low-income contexts. The Broad Institute constructed their licensing model for CRISPR technology for non-commercial research and agriculture uses, while limiting the license from applications of gene drives or human germline editing, as an example[14].

---

[12] Inventorship Guidance for AI-Assisted Inventions, *supra* note 3.
[13] Rebecca S. Eisenberg, *Patents and the Progress of Science: Exclusive Rights and Experimental Use*, 56 THE UNIVERSITY OF CHICAGO LAW REVIEW 1017 (1989), https://www.jstor.org/stable/1599761?origin=crossref (last visited Apr 24, 2025).

These perspectives reflect a normative information practice—one that sees patents as something more than legal or technical documents, but artifacts that have social and moral responsibilities associated with them. From the library and information science (LIS) perspective ethical licensing falls within an open science/open access framework that embraces equity in information acquisition and dissemination.

Still, ethical licensing leads to complications. Ethical licensing adds another layer of complexity to the legal metadata of patents, and such complexity adds hurdles to potential discoverability, interoperability, and enforceability. When ethical constraints are encoded into formal systems, information and professionals, and policy designers are weighing considerations for transparency, accessibility, and regulation.

### 4.3 Indigenous Knowledge and Biopiracy

The issue of patenting traditional or indigenous knowledge is particularly controversial. Examples of biopiracy—where companies have patented compounds or practices used by indigenous communities for centuries—indicate the inadequacy of current information systems to accommodate non-Western knowledge systems. Databases such as India's Traditional Knowledge Digital Library (TKDL) have developed for the purpose can documenting this knowledge and preventing its misappropriation[15].

In this regard, utilizing information science to bridge epistemic diversity entails designing systems that can capture non-codified or culturally situated knowledge, safeguard it, and fairly represent it. It involves rethinking taxonomies, ontologies, and even the words we use to classify patent categories.

## 5. Implications for Information Professionals and Policy

The changing geographies of patent law with respect to AI, biotechnology, and international competition for innovation, offers both challenges and opportunities to information professionals. As guardians of knowledge infrastructures, librarians, data curators, and information scientists have a unique opportunity to change how innovation is documented, accessed, and ethically governed.

### 5.1 Evolving Roles in Patent Information Management

Information professionals are being asked to have insight into and deal with patent metadata, classification, and indexing systems more frequently. With complexity in claiming AI-derived content and biotech inventions, cataloguing and indexing skills are now required to extend into interpreting computational creativity, tracing the provenance of inventorship, and understanding new ethical metadata (e.g., usage limitations or access terms).

---

[14] Information about licensing CRISPR systems, including for clinical use, @BROADINSTITUTE (2014), https://www.broadinstitute.org/partnerships/office-strategic-alliances-and-partnering/information-about-licensing-crispr-genome-edi (last visited Apr 24, 2025).

[15] Measures taken to protect the ancient and traditional knowledge of the country, https://pib.gov.in/pib.gov.in/Pressreleaseshare.aspx?PRID=1908022 (last visited Apr 24, 2025).

Patent information systems will need to change to:

- Disclose the role of AI in metadata fields.
- Record ethical restrictions in licensing terms.
- Embed Indigenous and traditional knowledge using culturally appropriate taxonomy.

This changes the role of the librarian from caretaker to the architect of innovation ecosystems, designing databases and interfaces that facilitate equitable knowledge flows.

### 5.2 Ethical Stewardship and Knowledge Justice

Information professionals have historically acted ethically as stewards of access. With patenting increasingly involved with commodifying critical technologies (e.g., medicines, climate mitigation tools), the professional responsibilities extend beyond stewardship to "knowledge justice", i.e., the fair distribution and representation of knowledge across socio-economic, geographic, and epistemic divides.

Some actions involve:

- supporting open patent databases as well as public domain databases,
- Encouraging ethical licensing via the policy consultation process,
- Working with policymakers to create inclusive frameworks for knowledge accessibility.

WIPO's Technology and Innovation Support Centres (TISCs) and India's Traditional Knowledge Digital Library (TKDL) are just two well-known examples of how LIS frameworks can protect information from exclusion, appropriation, or invisibility in the global knowledge economy[16].

### 5.3 Policy Input and Interdisciplinary Collaboration

While patent systems are lagging behind transformations brought about by technology, there is greater reliance upon interdisciplinary expertise by policy makers to develop flexible regulations. Information professionals can assist in this process by providing evidence-based information about:

- Effectiveness of information retrieval for AI-generated inventions.
- Accessibility implications of patent protection for disadvantaged populations.
- Best practices for data ethics, openness, and archival integrity.

By being engaged in legal policy reform and standardization activities (e.g., metadata schemas, open access mandates), information professionals are policy shapers as much as they are data custodians.

---

[16] Technology and Innovation Support Centers, TISC, https://www.wipo.int/web/tisc (last visited Apr 24, 2025).

With information being both power and property, the ethical and strategic management of patent data is now a public service, not just a technical function. Information professionals have the systems, frameworks, and ethical duty to steward patent data in a way that ensures knowledge created through innovation systems is accessible, traceable, and accountable to society. Information professionals are no longer in the ancillary role—they are integral to the future of equitable innovation.

6. **Conclusion**

As discussed in this section, the realm of patents—top of mind for lawmakers and economists for a long period—is presently undergoing significant changes. New technologies related to artificial intelligence and biotechnology are altering not only what is capable of being invented but they are also creating tensions with fundamental assumptions around who invents, who owns, and who can use the outcomes of innovation. It is not mere curiosity that creates challenges for innovators but rather a deep informational crisis and thus demands a serious response from the field of information science once again!

When patents are viewed as knowledge artifacts, an emphasis on knowledge as practice emerges and not just as ideas enshrined in legal documents. This realization leads us deeper into the connections we have drawn between intellectual property regimes and scholarly communication infrastructures, ethical stewardship, and access equity.

Major take-aways from this chapter include:

- A lot of AI generated discoveries open up epistemological gaps in the attribution and traceability of knowledge, which may need new metadata and ethical disclosure.
- Biotechnology patents pose urgent ethical-information dilemmas around commodification, traditional knowledge, and inclusive documentation.
- Global patent competition emphasizes transparent, interoperable and socially responsible information systems.
- Information professionals can create pro-active interdisciplinary roles as ethical stewards of knowledge, database designers, and policy advocates to facilitate fair innovation ecosystems.

Overall, the future of innovation governance must be as much about information ethics, knowledge organization, and access design as it is about law or economics. Patents are more than just about protecting inventions; they are about how knowledge circulates in society. This gives information science an important part of the role of facilitating an equitable, inclusive and intelligible global innovation system.